\documentclass[
reprint,
superscriptaddress,
amsmath,amssymb,
aps,
]{revtex4-2}

\usepackage{graphicx}
\graphicspath{{./Figure}}
\usepackage{dcolumn} 
\usepackage{bm} 
\usepackage{braket}
\usepackage{xcolor}
\usepackage{dsfont}
\usepackage{float}

\DeclareMathOperator{\Tr}{Tr}

\newcommand{\mt}[1]{\mathcal{#1}}
\newcommand{\al}{\alpha}

\newcommand{\bt}{\beta}

\newcommand{\Gm}{\Gamma}
\newcommand{\lmb}{\lambda}
\newcommand{\w}{\omega}
\newcommand{\W}{\Omega}
\newcommand{\s}{\sigma}
\newcommand{\Sg}{\Sigma}
\newcommand{\ur}{\uparrow}
\newcommand{\dg}{\dagger}
\newcommand{\rr}{\rightarrow}
\newcommand{\dr}{\downarrow}
\newcommand{\mi}{\mathrm{i}}

\newcommand{\gad}{\lambda_{_{(+)}}}
\usepackage{comment}

\begin{document}

\preprint{APS/123-QED}

\title{Singlet, triplet, and mixed all-to-all pairing states emerging from incoherent fermions} 

\author{Jagannath Sutradhar}
\affiliation{Physics Department, Ariel University, Ariel 40700, Israel}
\affiliation{Department of Physics, Bar Ilan University, Ramat Gan 5290002, Israel}
\email{sutradj@biu.ac.il}
\author{Jonathan Ruhman}%
\affiliation{Department of Physics, Bar Ilan University, Ramat Gan 5290002, Israel}
\email{jonathan.ruhman@biu.ac.il}
\author{Avraham Klein}
\affiliation{Physics Department, Ariel University, Ariel 40700, Israel}

\date{\today}

\begin{abstract}

The electron-electron and electron-phonon coupling in complex materials can be more complicated than simple density-density interactions, involving intertwined dynamics of spin, charge, and spatial symmetries. This motivates studying universal models with complex interactions, and studying whether in this case BCS-type singlet pairing is still the ``natural'' fate of the system. To this end, we construct a Yukawa-SYK model with nonlocal couplings in both spin and charge channels. Furthermore, we provide for time-reversal-symmetry breaking dynamics by averaging over the Gaussian Unitary ensemble rather than the Orthogonal ensemble. We find that the ground state of the system can be an orbitally nonlocal superconducting state arising from incoherent fermions with no BCS-like analog.  
The superconductivity has an equal tendency to triplet and singlet pairing states separated by a non-Fermi liquid phase. We further study the fate of the system within the superconducting phase and find that the expected ground state, away from the critical point, is a mixed singlet/triplet state. Finally, we find that while at $T_c$ the triplet and singlet transitions are dual to one another, below $T_c$ the duality is broken, with the triplet state more susceptible to orbital fluctuations just by virtue of its symmetry. Our results indicate that such fluctuation-induced mixed states may be an inherent feature of strongly correlated materials.
\end{abstract}

\maketitle


\textit{Introduction--}
One of the best-known avenues to analyze superconductivity beyond the classic  Bardeen-Cooper-Schrieffer (BCS) paradigm of weakly coupled Fermi liquids, is studying toy models of fermions coupled to a soft boson~\cite{Millis1993_PRB, Abanov_2000, Wang2001_PRL, Abanov_2001, Abanov_2003, ChubukovScmalian2005, PhysRevLett.117.157001, Metlitski2015_PRB, Raghu2015_QCP, Lederer2015_NemQCP, Gaopei2021_YSYK, Wonjune2022_YSYK, Andrew2021, WangWei2021_YSYK, Valetinis2023_BCStoYSYK}. The boson encodes both the dynamics and the symmetry properties of the interaction. Of particular focus has been the fate of such models when approaching a quantum critical point (QCP), such that the diverging correlation length creates both long-ranged and strong pairing interactions, and drives the fermions incoherent.
An overwhelming majority of studies consider \emph{spin-singlet} pairing \cite{Gaopei2021_YSYK, Wonjune2022_YSYK, Andrew2021, WangWei2021_YSYK, Valetinis2023_BCStoYSYK}. This is probably because most known superconductors are singlets, and the most common pairing mediators with the simplest symmetry properties (e.g. phonons) tend to prefer singlet pairing~\cite{brydon2014odd}.
\begin{figure}
   \centering
    \includegraphics[width=\hsize]{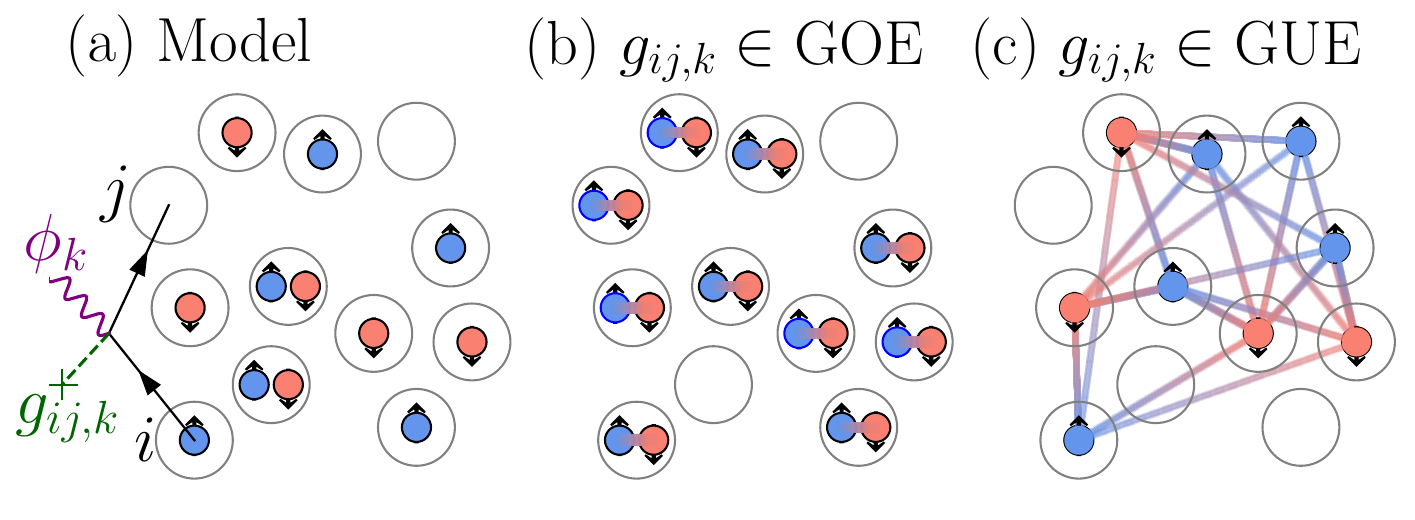}
  \includegraphics[width=0.7\hsize]{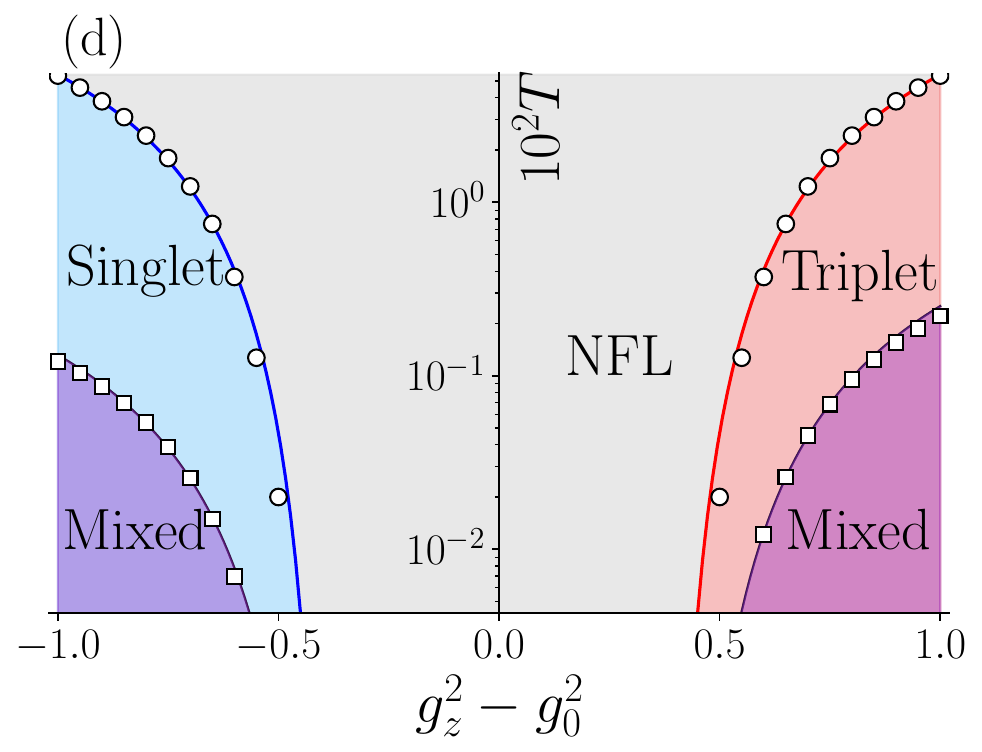}
    \caption{\textbf{(a)} Schematic of the model Hamiltonian [Eq.~\eqref{Eq:Ham}]: $N$ spinful Fermionic orbitals (circles) coupled to $M$ bosons (wiggly line) via random coupling $g_{ij,k}$, scattering electrons between orbitals $i$ and $j$. The scattering occurs in both density ($\s^0$) and spin ($\s^z$) channels with variances $g_0^2$ and $g_z^2$.
\textbf{(b)} Earlier studies averaged $g_{ij,k}$ over the GOE, leading to intra-orbital singlet pairing.
\textbf{(c)} In our study, $g_{ij,k}$ is drawn from the GUE, resulting in an ``all-to-all" inter-orbital pairing between all possible up-down spin pairs. \textbf{(d)} The phase diagram showing dual triplet and singlet SC phases, terminating at QCPs and separated by an NFL phase. At $T < T_c$, both singlet and triplet phases can induce a coexisting singlet/triplet state,
with the SC ground state transiting to a mixed pairing state.
The dots represent a numerical solution of the 
gap equation and the lines depict a fit to our analytical results, see the text for further details.}
    \label{Fig:summary}
\end{figure}

In truth, though, 
sufficiently complex materials should host interactions 
with intertwined attraction and repulsion, involving both charge and spin, as well as fluctuations that break lattice, time-reversal, and inversion symmetries. Such interactions have been considered for a host of candidate materials~\cite{Gorkov2001_PRL, Bauer2004_PRL, Frigeri2009_SC_noInversion, Vorontsov2009_FM_SC, Chubukov2003_FM_SC, Rech2006_PRB, Bert2011_imageFeSC,  Chubukov2012_AnRev, Scalapino2012_RMP, Hinojosa2014_TRS_SC, Klein2018_NemQCP, Chubukov2022_SC_rep, Zhaou2022_spin_polarizedSC, Esterlis_2019}. In addition, a variety of superconducting systems appear to evince phenomena that are neither clearly singlet-like nor triplet-like, raising the question of how likely mixed singlet/triplet states are in these systems~\cite{Frigeri2009_SC_noInversion, pustogow2019constraints, ribak2020chiral, salmani2020polar, hayes2021multicomponent}. This motivates finding a universal model of strongly correlated pairing that is ``agnostic'' as regards spin singlet and triplet pairing, as well as various spacetime and lattice symmetries. With such a model one can study the properties of strongly correlated superconductors, investigate whether spin singlet and triplet pairing have equal footing or not, and look for exotic pairing states that might be missing in ``garden-variety'' models.

In this Letter, we develop and study such a model. Our starting point is a Yukawa-Sachdev-Ye-Kitaev (Y-SYK) model ~\cite{Esterlis_2019, wang2020, Andrew2021}, where $N$ fermions are randomly coupled to $M$ bosons in 0+1D, as shown in Fig.~\ref{Fig:summary}(a).  This model is analytically solvable in the large-$N$ large-$M$ limit and exhibits both non-Fermi liquid (NFL) behavior and high-temperature superconductivity. 
We consider both spin and charge coupling, with independent coupling constants, $g_z$ and $g_0$, respectively. Moreover, we allow for time-reversal-symmetry (TRS) breaking fluctuations by drawing the coupling matrix from the Gaussian Unitary Ensemble (GUE), rather than the more usual Orthogonal Ensemble (GOE). 
 
 We find that upon lowering the temperature the bosons become critical and the NFL becomes unstable to an \emph{inter-orbital} pairing state $\Phi \sim \langle c_i c_j \rangle, i\neq j$ [Fig.~\ref{Fig:summary}(c)], where $i,j$ denote orbitals and we suppressed spin indices. Such pairing is qualitatively distinct from the ``standard'' BCS type $\Phi \sim \langle c_i c_i \rangle$ that arises in the GOE averaging [Fig.~\ref{Fig:summary}(b)].  
Furthermore, this nonlocal state is tuned between triplet and singlet pairing depending on the relative coupling strengths as shown in the phase diagram Fig.~\ref{Fig:summary}(d). The two states are separated by an NFL phase. We also study the pairing state below $T_c$. We find that the low-$T$ pairing state is generically a nonlocal triplet-singlet mixture. 
 Our results show that the previously deemed 
 NFL phase can become unstable to pairing once the non-local state is taken into account. This includes models where the random couplings are a sum of two random numbers drawn from the GOE and GUE~\cite{Andrew2021}.

Thus, systems with both charge and spin fluctuations have a rich landscape of superconducting phases that go beyond the usual paradigms of unconventional superconductivity and can involve not just broken symmetries such as time reversal, but also spatially nonlocal correlations even at the mean-field level. In these systems, an introduction of nonlinear contributions plays a crucial role in shaping the phase diagram, even in the absence of static symmetry breaking. We describe our model and results in detail below.

\textit{Model--}
We consider $M$ bosonic fields, $\phi_k$,  randomly coupled to $N$  fermions $c_{i\al}$ as schematically depicted in Fig.~\ref{Fig:summary}(a) and expressed as
\begin{align}
H &= H_z + H_0\ + \sum_{k} (\pi_{k}^{2} + m^{2}_{0}\phi_{k}^2) \,, \label{Eq:Ham}
  \\
H_{a} &= 1/\sqrt{MN}\sum_{ijk\al\bt} g^{a}_{ij,k} \phi_k c_{i\al}^\dg \s^{a}_{\al\bt} c_{j\bt} \quad(a=z,0),\nonumber
\end{align}
where $\pi_k$ are the momenta fields conjugate to $\phi_k$, and $\alpha,\beta$ are spin indices. 
The random couplings $g^a_{ij,k}$ are completely uncorrelated between the spin and charge components, denoted by $a=z,0$
respectively. The interaction form is chosen to allow for universal spin and charge fluctuations, while avoiding stability issues that arise from SU(2) spin fluctuations~\cite{Chubukov2003_PRL, Chubukov2009_PRL}.
The couplings are drawn from the GUE, such that the average of the real and imaginary parts:
$\overline{\mathrm{Re}(g^{a}_{ij,k})\mathrm{Re}(g^{a}_{i'j',k'}})=g_a^2 \delta_{kk'}(\delta_{ii'}\delta_{jj'}+\delta_{ij'}\delta_{ji'})$ and 
$\overline{\mathrm{Im}(g^{a}_{ijk})\mathrm{Im}(g^{a}_{i'j'k'})}=g_a^2 \delta_{kk'}(\delta_{ii'}\delta_{jj'}-\delta_{ij'}\delta_{ji'})$, respectively and $g_a$ is the coupling strength.
Note that 
the complex $g^{a}_{ij,k}$ 
manifestly breaks TRS, which is restored upon disorder-averaging.
Previous studies considered a similar model with TRS by drawing the coupling from the GOE. In that case, there is a large-$N$ instability towards an intra-orbital singlet pairing phase characterized by the non-vanishing expectation value
$\sum_i \langle c_{i\uparrow}^\dg c_{i\downarrow}^\dg\rangle \ne0$.
In contrast, the GUE-averaged action avoids the intra-orbital pairing and opens a path to more exotic pairing states. The distinct pairing forms are illustrated in Figs.~\ref{Fig:summary}(b) and (c). 

\textit{The triplet normal state and pairing instability--}
Before jumping into the full Hamiltonian in Eq.~\eqref{Eq:Ham}, we
start by focusing on the spin fluctuation $H_z$, which governs the triplet pairing and set 
$g_0 = 0$.
We then show that 
the complimentary singlet pairing scenario where 
$g_z = 0, g_0 \neq 0$ can be obtained directly from the triplet case. 
Finally we study the interplay between both spin and charge couplings.

\begin{figure}
\centering
    \includegraphics[width=0.9\linewidth]{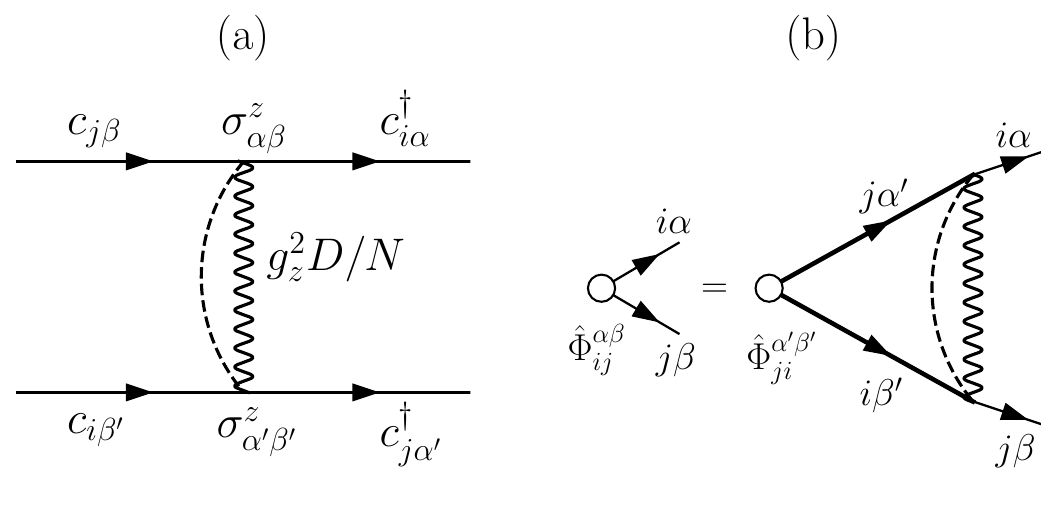}
    \caption{ \textbf{Feynman diagrams:} \textbf{(a)} Schematic of the interacting four-fermionic term mediated by the bosonic propagator $D$ (wavy line) after disorder average represented by the dashed line. 
    \textbf{(b)} The linearized gap equation. Note that the exchange in $ij$ indices in $\hat\Phi$ on the right side makes one of the triplet channels attractive.
    }
    \label{Fig:Feynman_diagram}
\end{figure}

As usual, we disorder-average the partition function associated with the Hamiltonian in Eq.~\eqref{Eq:Ham} with $H_0=0$ using the replica trick and consider only the replica-diagonal term in the action \cite{Esterlis_2019}. The form of the interacting part of the action is 
\begin{equation}\label{eq:S_I}
    \mt{S}_I \sim -g_z^2 \sum_{k,i,..} \int d\tau \phi_k c_{i\al}^\dg \s^z_{\al\bt} c_{j\bt} \int d\tau^\prime \phi_k c_{j\al^\prime}^\dg \s^z_{\al^\prime\bt^\prime} c_{i\bt^\prime}\,,
\end{equation} (see the supplemental materials [SM] for details),
and the corresponding Feynman diagram is shown in Fig.~\ref{Fig:Feynman_diagram}(a). 
Superficially, Eq.~\eqref{eq:S_I} appears similar to ``standard'' spin-fermion interactions whose QC dynamics have been studied extensively~\cite{Gaopei2021_YSYK, Wonjune2022_YSYK, Andrew2021, WangWei2021_YSYK, Valetinis2023_BCStoYSYK, Chubukov_2020_gammaModel,Chubukov_2021}. However, the interaction in $H_z$ of Eq.~\eqref{Eq:Ham} involves both spin-flips \emph{and} nonlocality, very different from the usual onsite magnetic interactions, giving rise to some nontrivial physics.
In what follows we study the fate of the fermions in the presence of this interaction.

Connecting different legs in Fig.~\ref{Fig:Feynman_diagram}(a) we obtain the self-energies in the large $N,M$ limit. 
The Green's function is diagonal both in orbital and spin basis~[SM] and it is defined as $G(\w_n)=1/[\mi \w_n+\mi \Sg(\w_n)]$.
At finite $T$, the electronic self-energy is given by 
\begin{equation}
\Sigma(\w_n)=\mi g^{2}_z T \sum_{m} D(\w_n-\w_m) \dfrac{1}{2} \Tr \big[ \s^{z} G(\w_m)\s^z \big] \,,
\label{Eq:Sigma_diagram}
\end{equation}
where $\w_n=(2n+1)\pi T$, $n\in \mathbb Z$ are fermionic Matsubara frequencies and Tr represents trace over the spin indices. 
 $D(\W_n)=[\W_n^2 + m_0^2 + \Pi(\W_n)]^{-1}$ is the bosonic propagator. The bosonic self-energy is given by   
\begin{equation}
\Pi(\W_n)=2g^{2}_z T \dfrac{N}{M} \sum_{m} \Tr\big[\s^{z} G(\w_m+\W_n) \s^z G(\w_m) \big] \,,
\label{Eq:Pi_diagram}
\end{equation}
where $\W_n=2n\pi T$ are the bosonic Matsubara frequencies. 
Similar to what occurs in previously studied models
~\cite{wang2020,Andrew2021}, Eqs.~\eqref{Eq:Sigma_diagram} and \eqref{Eq:Pi_diagram} give rise to NFL behavior, with 
$ \Sg(\w)\sim \mathrm{sgn}(\w) |\w|^{(1-\eta)/2} 
$ and
$\Pi(\W)-\Pi(0) \sim g^{2} |\W|^{\eta}
$ at low $T,\w$, and self-tuned criticality that renormalizes the boson mass to exactly zero.
The exponent $\eta$ varies from $0$ to $1$ as a function of $N/M$ [SM].

The form of the interaction naturally leads us to consider an inter-orbital pairing state $\Phi_{ij}^{\al\bt}=\braket{c_{i\al}^\dg c_{j\bt}^\dg}$, which includes all possible orbital pairs. A somewhat similar type of pairing was considered in 
Ref.~\cite{wang2020}.
The linearized equation for the inter-orbital  order parameter, which corresponds to Fig.~\ref{Fig:Feynman_diagram}(b), is given by
\begin{flalign}
    \hat{\Phi}^{\al\bt}_{ij}(\w_n)=\dfrac{g^2_z}{N} T \sum_{m ,\al^\prime, \bt^\prime}  D(\w_n-\w_m) \s^z_{\al\al'} \qquad\qquad
    \nonumber\\
   G(\w_m) \hat\Phi^{\al'\bt'}_{ji}(\w_m) G(-\w_m) \s^z_{\bt\bt'} \,.
    \label{Eq:gap_fn_in_detail}
\end{flalign}
Note the orbital exchange $i\leftrightarrow j$ between the left and right-hand side of the equation, which is a consequence of the nonlocality discussed above. 
The antisymmetry of the pairing function is enforced by the orbital exchange, i.e., $\hat\Phi_{ij}^{\al\bt}=-\hat\Phi_{ji}^{\al\bt}$. Consequently, Eq.~\eqref{Eq:gap_fn_in_detail} indicates attraction only when $\hat{\Phi}^{\al\bt}_{ij}=\hat{\Phi}_{ij} \s^x_{\al\bt}$, which signifies a triplet pairing state. 
Note that this is the \emph{only} attractive channel, with the other two spin channels, as well as the singlet channel, being repulsive. 
This is in stark contrast to what occurs in quantum critical models with a local spin-fermion interaction, where each magnetic channel is attractive in two spin channels and repulsive in the third~\cite{Chubukov2003_PRL}.
To simplify 
Eq.~\eqref{Eq:gap_fn_in_detail}, we neglect the orbital phase fluctuation and consider $\hat\Phi_{ij}=\Phi_t A_{ij}$, where $\Phi_t$ is the amplitude of the pairing function dependent only on frequency, and $A_{ij}$ is a skew-symmetric matrix, such that all elements $A_{i>j} = 1, A_{i<j} = -1, A_{i=j} = 0$.
Choosing a different antisymmetric matrix $A_{ij}$ yields the same results [SM]. 
We will discuss the origin and consequence of this degeneracy later on. 

 At $T=0$ Eq.~\eqref{Eq:gap_fn_in_detail} has  a SC instability in the regime $\sqrt{2M}>N$ and in the large $M$ and $N$ limit~\cite{wang2020,Andrew2021}. In contrast to the BCS theory, for $\sqrt{2M}<N$ even strong attraction does not lead to SC. 
At finite $T$, the bosons develop a mass correction, and the pairing equation resembles that in Ref.~\cite{Andrew2021}. 
The critical temperature ($T_c$) for the phase transition from the NFL to SC follows $T_c\sim g^2_z$ [SM]. 

\textit{The singlet channel and mixed case--} We next consider
the pure singlet case, $g_z = 0,\ g_0 \neq 0$,
the gap equation exhibits attraction in the singlet channel, characterized by the pairing function $\hat{\Phi}_{ij}^{\al\bt}=\mi \Phi_s S_{ij} \s^y_{\al\bt}$,
where $S$ is a symmetric traceless matrix. 
As before, we neglect the fluctuations and set all the off-diagonal elements of the 
matrix 
to be $1$, $S_{ij} = 1-\delta_{ij}$
(for further discussion see SM). The normal state and SC gap equations remain the same as in the triplet case. Consequently, the phase diagram resembles that in the earlier case, replacing the triplet SC with a singlet SC. 
Thus, above the critical $\sqrt{M}/N$ ratio, the ground state for our model with either of the couplings is always SC arising from within the incoherent NFL state. This is again in contrast with the case of  GOE averaging, where the BCS-type local pairing sets in at the same energy scale as NFL physics~\cite{Esterlis_2019, Andrew2021}.
For the singlet case, a BCS-type local pairing $\hat{\Phi}^{\alpha\beta}_{ij} \propto \delta_{ij}i\sigma^y_{\alpha\beta}$ is degenerate with the nonlocal pairing. For simplicity we consider henceforth only the nonlocal state, since it keeps the physics qualitatively the same [SM].

We are now ready to consider the full Hamiltonian of Eq.~\eqref{Eq:Ham}. 
As $g^0_{ij,k}$ and $g^z_{ij,k}$ are uncorrelated, the interacting part of the disorder averaged action is a sum of two terms similar to Fig.~\ref{Fig:Feynman_diagram}(a): one for $\s^z$ as earlier and another for $\s^0$, with their respective strengths $g_z$ and $g_0$. The normal state solution remains NFL.
Within the linearized gap equation, there is a critical strength $\lmb_c$ of $\lmb=|g_z^2-g_0^2|$ beyond which the SC phase emerges. The state is either singlet (for $g_0>g_z$) or triplet (for $g_z > g_0$), with no mixing [see Fig.~\ref{Fig:summary}(d)]. The critical temperature follows $T_c\sim \mathrm{exp}(-\pi/\sqrt{\lmb-\lmb_c})$ \cite{Andrew2021}[SM].
Between the SC phases, the system remains an NFL down to $T=0$.

\begin{figure}
    \centering
    \includegraphics[width=0.9\linewidth]{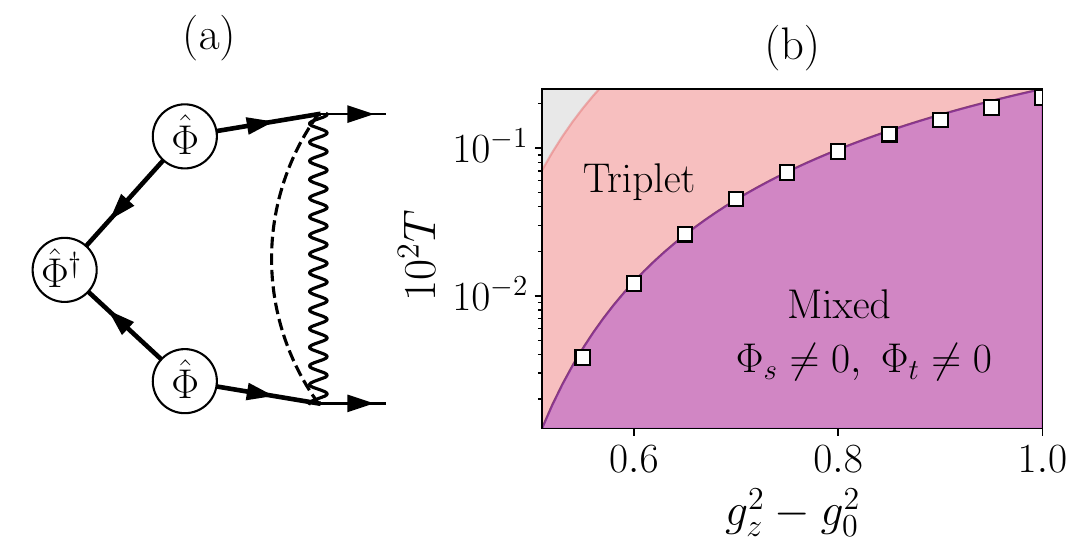}
    \caption{\textbf{Nonlinearity in the gap equation:} \textbf{(a)} The Feynman diagram represents the first nonlinear term in the expansion of the gap equation. \textbf{(b)} Variation of the critical temperature $T_{t\rr m}$ to the mixed state as a function of
    $g_z^2-g_0^2$. This figure zooms at the segment of Fig.~\ref{Fig:summary}(d) depicting the transition from the triplet to the mixed state. The squares denote the numerical data points and the dashed line represents a fitting to the curve $T_{t\rr m}\sim  \mathrm{exp}\big(-\pi(4/(3+\eta))/\sqrt{\lmb-\lmb_c}\big)$ [SM] for $g_0^2=1$.}
    \label{Fig:nonlin}
\end{figure}

\textit{Solution of the gap equation below $T_c$--}
To explore the phase diagram at $T<T_c$
we must turn to the nonlinear gap equation. 
To begin the analysis, let us consider the case with only one finite coupling, $\s^z$ or $\s^0$. 
Obtaining an analytic solution to the full non-linear equations is challenging due to the form of the matrices $A$ and $S$, which couple all orbitals. However, near $T_c$, we can study the first non-linear correction to the linearized gap equation by dressing one of the Green's functions in the diagram Fig.~\ref{Fig:Feynman_diagram}(b)  with a correction of order $\sim |\Phi_{t(s)}(\w_m)|^2 \Phi_{t(s)}(\w_m) $, as depicted in Fig.~\ref{Fig:nonlin}(a) [SM]. 
The nonlinear term has an opposite sign compared to the linear term [Eq.~\eqref{Eq:Phits_nonlin}] and gives rise to the usual $\Phi_a \sim \sqrt{T_c-T}$ behavior.

We now turn to the case involving both $\s^z$ and $\s^0$ couplings. We introduce a \emph{mixed} pairing function $\hat\Phi=\Phi_t A \s^x+\mi\Phi_s S \s^y$. 
As mentioned earlier, in the linear approximation, $\Phi_{t}$ and $\Phi_{s}$ do not interact. However, higher-order terms introduce an interaction between $\Phi_{t}$ and $\Phi_{s}$. The equation for $\Phi_{t(s)}$ can be written in simplified form as, 
\begin{widetext}
\begin{equation}
\Phi_{t(s)}(\w_n)\approx
\big[ g_{z(0)}^2-g_{0(z)}^2 \big] \dfrac{T}{N} \sum_{m} \dfrac{D(\w_n-\w_m)}{|\w_m+\Sg(\w_m)|^{2}} \Bigg[ 
\Phi_{t(s)}(\w_m) - \dfrac{\Phi_{t(s)}(\w_m)}{|\w_m+\Sg(\w_m)|^{2}} 
\Big\{ \zeta_{t(s)} \Phi_{t(s)}^2(\w_m)  + \dfrac{2}{3}\Phi^2_{s(t)}(\w_m)  \Big\}
\Bigg] \,,
\label{Eq:Phits_nonlin}
\end{equation}
\end{widetext}
where
we rescaled $\Phi_{t(s)}\rightarrow \Phi_{t(s)}/N$ and neglected $1/N^2$ terms. The factor $\zeta_t=1/3$ and $\zeta_s=1$. We also took both fields to be real, since their phase turns out to be locked by the nonlinear terms (see SM for the full equation and analysis).

The non-linear term implies the emergence of a mixed pairing state at lower temperatures. 
For example, let us consider $g_z^2>g_0^2$ and $T$ is just below $T_c$, which results in $\Phi_t\ne 0$ and $\Phi_s=0$, i.e., the SC phase consists purely of triplet pairing. 
Now, in the equation for $\Phi_s$ the nonlinear contribution $\propto \Phi_t^2 \Phi_s$ is attractive although the linear term is repulsive. 
Consequently, as we lower the temperature and $|\Phi_t|$ increases, the attraction at low frequencies 
dominates over the linear repulsive term, triggering a second instability below some critical $T$.
Thus, the SC phase becomes a superposition of singlet and triplet pairing. 
This mechanism is reminiscent of the competition between repulsive and attractive interactions in the Anderson-Morel model~\cite{AndersonMorel_62}.
The arguments for $g_0^2 > g_z^2$ are exactly analogous, and this gives rise to the phase diagram plotted in Fig.~\ref{Fig:summary}(d) for $\eta=0.68$ and $N=1$. We confirm our analysis by numerically determining the critical temperature $T_{t(s)\rr m}$ [SM]. 
An analysis of the nonlinear gap equation within the SC phase reveals that the mixed state is stable all the way to the dominant order's critical point at $\lambda=\lambda_c$, and our numerics confirm this analysis. This contradicts the decoupling of the two states that occurs at the level of the linearized gap equation. Thus, the QCP appears to be a singular point with different behavior depending on how it is approached in the phase diagram. This is probably because the Ginzburg-Landau expansion has coefficients that diverge at low temperatures. To conclusively determine whether the QCPs to the single state and mixed state converge or not requires an analysis of the full gap equation, which is beyond the scope of this work.

The phase diagram Fig.~\ref{Fig:summary}(d) implies a duality between the singlet and triplet states.
However, an interesting distinction exists between them. To understand it, we recall that the $T_c$ obtained in the triplet state does not depend on the choice of the antisymmetric matrix $A$. The existence of a degeneracy is a manifestation of an $S_N$ permutation symmetry of the disorder-averaged action Eq.~\eqref{eq:S_I}. On the contrary, no such degeneracy exists in the singlet case, such that $S$ is unique. Indeed, the matrix $S$ forms a \emph{one-dimensional} irreducible representation (irrep) of $S_N$, while $A$ belongs to a multidimensional irrep, with $N(N-1)/2$ different elements. 
Nonetheless, it should be noted that in any case the gap between the different irreps vanishes in the $N\to \infty$ limit. 

\textit{Discussion--}
We have studied the Yukawa-SYK model for a random coupling between $M$ bosons and $N$ fermions drawn from the GUE, which corresponds to the limit where time-reversal symmetry is absent. We couple the bosons to both spin and charge. 
The intra-orbital pairing term vanishes upon averaging the random coupling (diagonal replica symmetry breaking). 
In the absence of this term, the leading pairing instability is a highly non-local inter-orbital pairing state with an equal superposition on all orbitals. 
By tuning the relative strength between the charge and spin couplings the ground state is tuned through a sequence of QCPs, separating between singlet, NFL and triplet phases. Going deeper into each one of these phases, there is an additional transition to a
singlet-triplet mixed state, as shown in the phase diagram Fig.~\ref{Fig:summary}(d).  

It is interesting to compare the  inter- and intra-orbital pairing order parameters with the standard BCS pairing wave function in a finite-dimensional system with translational invariance (i.e. with a Fermi surface). In the BCS wave function, the pair correlations are between $k$ and $-k$ and are therefore equivalent to intra-orbital pairing. On the other hand, the inter-orbital order parameter we presented here would imply pairing correlations between all momentum states. Thus, the wavefunction corresponding to the inter-orbital pairing state is inconsistent with the picture of momentum-space quasiparticles. In that sense, one may think of a GOE-averaged Y-SYK system as the quasiparticle weak coupling limit, and the GUE averaged system as a strong coupling limit. 
Moreover, the existence of an instability in the entire phase diagram implies that the NFL phase driven by TRS breaking is generically unstable to pairing in the Y-SYK model.

It is also interesting
that the triplet order parameter belongs to a \emph{multi-dimensional} irreducible representation of the permutation group, while the singlet belongs to a one-dimensional irrep. Consequently, we expect the triplet state to be more susceptible to quantum fluctuations, and thus the phase diagram in Fig.~\ref{Fig:summary}(d) will not retain the symmetry between the $g_z>g_0$ and $g_z<g_0$ limits where the transition into the triplet state will likely be suppressed.  

Finally, the Y-SYK action has a 1+1 AdS dual \cite{Hartnoll2008_PRL, Subir2015_PRX, schmalian2022holographicTriplet, Schmalian_2022_holography}, where $T_c$ is dual to an event horizon (indicating a black hole formation). Looking forward it is interesting to understand the gravity dual of the inter-orbital state found here. In particular, it may give access to black holes with finite angular momentum.  

\begin{acknowledgments}
We thank J. Schmalian, A.V. Chubukov, S.K. Saha, and H. Yerzhakov for interesting discussions. A.K. and J.R. acknowledge support by the Israel Science Foundation (ISF), and the Israeli Directorate for Defense Research and Development (DDR\&D) under grant No. 3467/21.
\end{acknowledgments}

\bibliography{reference}

\onecolumngrid
\begin{center}
     \Large {Supplementary Material}
 \end{center} 
\renewcommand\thesection{S\arabic{section}}
\renewcommand\theequation{S\arabic{equation}}
\renewcommand\thefigure{S\arabic{figure}} 

 \setcounter{section}{0}
 \setcounter{equation}{0}
 \setcounter{figure}{0}
 
In these supplementary material notes, we provide details on the calculations we referred to in the main text. We start with technical details regarding the action and the evaluation of the normal state properties. Most of these details can be found in previous works, but we include them for completeness. Then we give details on our solutions of both linear and nonlinear gap equations at finite $T$, and with both $g_z$ and $g_0$.
\section{Evaluation of the action}
 The partition function is defined as $Z=e^{-H/T}$ and the Hamiltonian $H$ is given in Eq.~(1) of the main text. We begin with deriving the action only for the spin channel, i.e. for $g_0=0$.  The disorder-averaged action is obtained using the replica trick. To do so, we first calculate $\braket{Z^{R}}_{\text{GUE}}$, where $R$ is an integer and $\braket{..}_{\text{GUE}}$ denotes average over the GUE disorder $g^z_{ij,k}$. The interacting part of the action is expressed as follows:
\begin{equation}
\mt{S}_I 
=
-g_z^{2} \int d\tau d\tau' \sum_{a,b=1}^{R} \sum_{k}\phi_{k,a}(\tau) \phi_{k,b}(\tau') \sum_{ij\al\bt\al'\bt'} c_{i\al, a}^\dg(\tau) \s^z_{\al\bt} c_{j\bt, a}(\tau) \ c^\dg_{j\al', b}(\tau') \s^z_{\al'\bt'} c_{i\bt',b}(\tau') \,.
\label{Eq:interaction_term1}
\end{equation}
Here $a,b=1,...,R$ signifies the replica indices. For simplification and further analysis, we assume the contribution in $\mt{S}_I$ arising only from the diagonal elements in the replica basis, allowing us to eliminate the replica structure. Consequently, the interacting part is given by
\begin{equation}
\mt{S}_I 
=
-g_z^{2} \int d\tau d\tau'  \sum_{k}\phi_{k}(\tau) \phi_{k}(\tau') \sum_{ij\al\bt\al'\bt'} c_{i\al}^\dg(\tau) \s^z_{\al\bt} c_{j\bt}(\tau) \ c^\dg_{j\al'}(\tau') \s^z_{\al'\bt'} c_{i\bt'}(\tau') \,, 
\end{equation}
which is represented using Feynman diagram in Fig.~2(a). In the presence of both spin and charge couplings, the interaction is expressed as
\begin{equation}
\mt{S}_I 
=
- \int d\tau d\tau'  \sum_{k}\phi_{k}(\tau) \phi_{k}(\tau') \sum_{ij\al\bt\al'\bt'} 
\sum_{a=0,z} g_a^{2} \ c_{i\al}^\dg(\tau) \s^a_{\al\bt} c_{j\bt}(\tau) \ c^\dg_{j\al'}(\tau') \s^a_{\al'\bt'} c_{i\bt'}(\tau') \,.
\label{SEq:Si_BothCoupling}
\end{equation}
 \section{Normal state solution}
The calculations shown in this section are only in the presence of the spin channel. 
From Fig.~2(a), it is evident that the self-energy is diagonal in the orbital basis. To obtain its spin structure let us consider a general self-energy $\mathbf{\Sg}_f =\Sg\s^0+ \vec{\Lambda}\cdot\vec{\s}$ incorporating all the spin components. The self-consistent self-energy equation is then given by
\[
\begin{split}
\mathbf{\Sg}_f(\w_n)
&=
\mi g^{2}_z T \sum_{m} D(\w_n-\w_m) \s^{z} [\w_m\s^0 +\mathbf{\Sg}_f(\w_m)]^{-1}\s^z  \,, \\
&= \mi g^{2}_z T \sum_{m} D(\w_n-\w_m) \s^{z} \dfrac{[\w_m + \Sg(\w_m)]\s^0-\vec{\Lambda}(\w_m)\cdot \vec{\s}} {[\w_m + \Sg(\w_m)]^2-|\vec{\Lambda}(\w_m)|^2} \s^z  \,.
\end{split}
\]
Due to the presence of the bare term $\w_m$ in the $\s^0$ component, $\Sg$ must be nonzero. This implies 
\begin{equation}
(\w_m + \Sg(\w_n))^2 > |\vec{\Lambda}|^2, \label{eq:sig0-domin}    
\end{equation} 
and consequently $\Lambda_z=0$, as there is a relative sign between the left and right side of the equation for the $z$-component. In the limit $\w\rightarrow 0$, there are two possibilities for $\Lambda_x$ and $\Lambda_y$. One possibility is that they decay with with a power-law form that is subleading compared to $\Sg$, which we can neglect. Alternatively we consider the following ansatz,
$(\Lambda_x,\Lambda_y)=\vec{Q}\Sg$, where $|\vec{Q}|<1$ is a constant due to the constraint of Eq. \eqref{eq:sig0-domin}. In this case,
\begin{align}
\vec{Q}\Sg(\w_n)
&= 
-\mi g^{2}_z T \sum_{m} D(\w_n-\w_m) \dfrac{\vec{Q} \Sg(\w_m)} {[\w_m + \Sg(\w_m)]^2-|\vec{\Lambda}(\w_m)|^2} \,, \\
\Rightarrow 
\vec{Q}\Sg(\w_n)
&=
\vec{Q}\Sg(\w_n)+\mi g^{2}_z T \sum_{m} D(\w_n-\w_m) \dfrac{\vec{Q} \w_m} {[\w_m + \Sg(\w_m)]^2-|\vec{\Lambda}(\w_m)|^2}  \,.
\end{align}
Therefore, $\vec{Q}=0$, leaving the self-energy contribution only from $\Sg\s^0$.

\begin{figure}[ht]
\centering
    \includegraphics[width=0.5\linewidth]{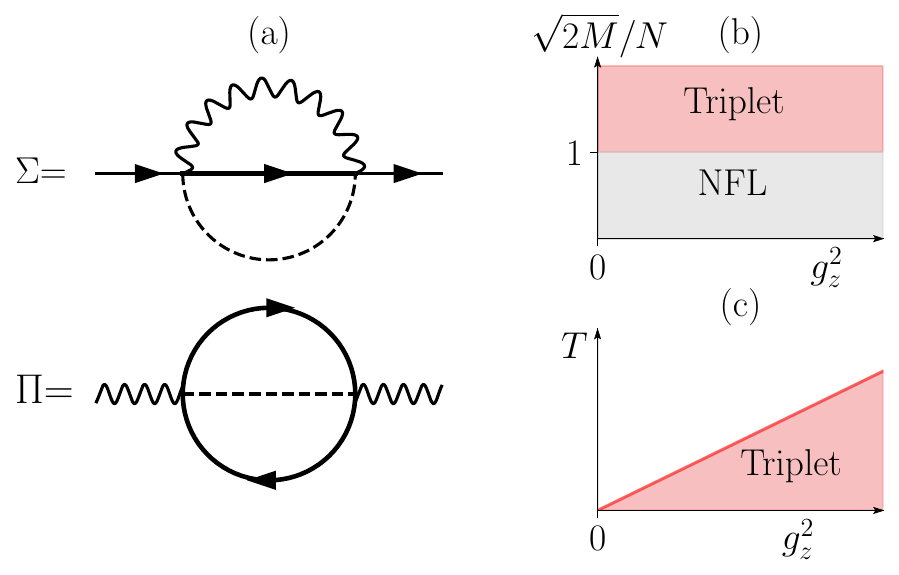}
    \caption{ \textbf{Linear gap equation:} \textbf{(a)} Feynman diagrams for evaluating the electronic self-energy (top) and bosonic self-energy (bottom). The bold lines signify the electronic Green's functions. 
    \textbf{(b)} Phase diagram at $T=0$. For large $M$ and $N$ when $\sqrt{2M}/N>1$, the system transits from NFL to the triplet SC phase. 
   \textbf{(c)} Phase diagram at finite temperature $T$. The transition temperature to the SC phase $T_c\sim g_z^2$.
    }
    \label{SFig:SelfEnergy_LinGap}
\end{figure}

\begin{figure}
\centering
\includegraphics[scale=0.5]{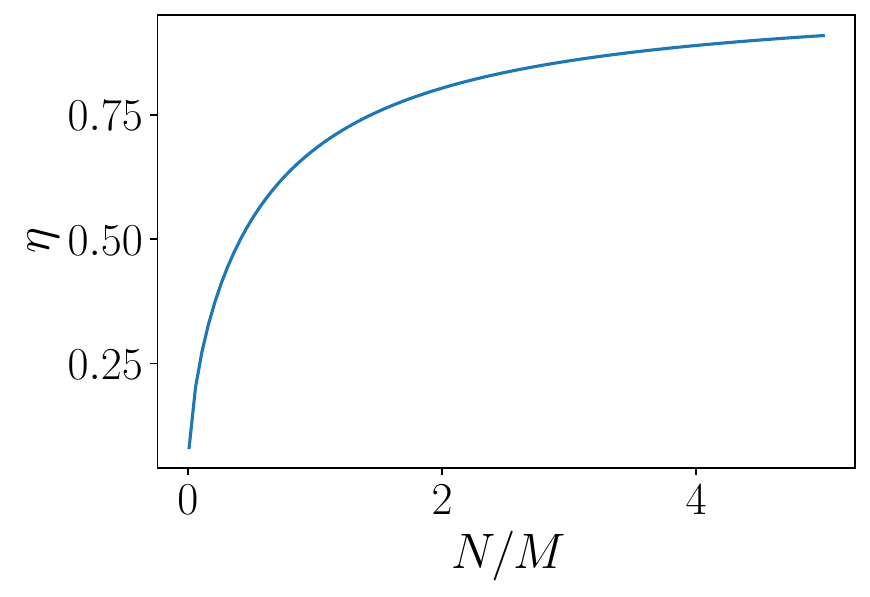}
\caption{$\eta$, which determines the power law of the self-energies, is shown as a function of $N/M$ using Eq.~\eqref{SEq:eta_NM}.}
\label{SFig:eta_vs_NM}
\end{figure}

 In the normal state $\Phi_t=0$. Here, we calculate the $T=0$ case. The self-consistent Eqs.~(3) and (4) are obtained using the Feynman diagram illustrated in Fig.~\ref{SFig:SelfEnergy_LinGap}(a). 
Following the approach in Ref.~\cite{Andrew2021}, we consider a power-law form for the electronic self-energy, namely
\begin{equation}
\Sg(\w)=\mathrm{sgn}(\w) a_{\eta}^{(1+\eta)/2}|\w|^{(1-\eta)/2} \,,
\end{equation}
i.e., the propagator is dominated by the self-energy at low energy, $\w\rightarrow 0$.
Consequently, the bosonic self-energy at low energy is given by
\begin{equation}
\begin{split}
\delta\Pi(\w)
&=
\Pi(\w)-\Pi(0) \,, \\
&=
-2g_z^{2}\dfrac{N}{M}\int \dfrac{d\w'}{2\pi} \Bigg[\dfrac{1}{\Sg(\w'+\w/2)\Sg(\w'-\w/2)}-\dfrac{1}{\Sg(\w')\Sg(\w')} \Bigg] \,, \\
&=
-2g_z^{2}\dfrac{N}{M} \dfrac{\Gamma^{2}(-x)}{2\Gamma(-2x)} \dfrac{1+\sec (\pi x)}{1/x-2} \dfrac{|\w|^{\eta}}{2\pi a_{\eta}^{1+\eta}} \,.
\end{split}
\label{SEq:bos_self_enT0}
\end{equation}
Here $x=(1-\eta)/2$. The condition required for critical boson, $\Pi(0)=-m_0^{2}$, yields
\[
a_{\eta}=
\dfrac{2}{\pi} \Big(\dfrac{g_z}{\w_0}\Big)^2 \dfrac{N}{M}\Gamma\Bigg(\dfrac{2\eta}{1+\eta} \Bigg) \Gamma\Bigg(\dfrac{3+\eta}{1+\eta} \Bigg) \,.
\]
The evaluation of $\Pi(0)$ requires the bare terms in $G(\w)$ to avoid divergence in the integration~\cite{Andrew2021}. We use the above expressions to calculate the electronic self-energy and verify the self-consistency:
\[
\begin{split}
&
\Sg(\w)= g_z^{2} \int \dfrac{d\w'}{2\pi} \dfrac{D(\w-\w')}{\w+\Sg(\w')} \,, \\
\Rightarrow
\Sg(\w)& =g_z^{2} \int \dfrac{d\w'}{2\pi} \dfrac{1}{\delta\Pi(\w-\w')\Sg(\w')} \,,  \ \ \text{for } \Sg(\w) \ \text{and} \ \delta\Pi(\w)\gg \w, \\
&=
\dfrac{M}{2N} \dfrac{1/x-2}{1+\sec(\pi x)} a_{\eta}^{(1+\eta)/2} \mathrm{sgn}(\w) |\w|^{(1-\eta)/2} \,. \\
\end{split}
\]
Self consistency of $\Sg(\w)$ relates $\eta$  with $N/M$ as
\begin{equation}
\dfrac{2N}{M}=\dfrac{1/x-2}{1+\sec(\pi x)}
=\dfrac{2\eta}{1-\eta} \dfrac{\tan(\pi\eta/2)}{\tan[\pi(1+\eta)/4]} \,,
\label{SEq:eta_NM}
\end{equation}
and plotted in Fig.~\ref{SFig:eta_vs_NM}. 
Using Eqs.~\eqref{SEq:bos_self_enT0} and \eqref{SEq:eta_NM}, we write 
$
\delta\Pi(\w)=g_z^2 b_{\eta}|\w|^{\eta}/a_{\eta}^{1+\eta} $, where
\[
b_{\eta}=-\dfrac{1}{4\pi}\dfrac{\Gm^2[(\eta-1)/2]}{\Gm(\eta-1)}.
\]
The expression of $b_\eta$ given above differs from that in Ref.~\cite{Andrew2021} because of a typo present in their expression.

The solution at $T>0$ is a more complicated version of the above, and we omit it for brevity as the details can be found in Ref.~\cite{Andrew2021}. 
We note that the normal state of our model has other regimes of behaviour, namely, a regime of free fermions and an impurity regime,~\cite{Andrew2021}  apart from the SYK regime considered here, depending on temperature, coupling strength, and the ratio of $N/M$.  However, we focus only on the SYK regime at a very low-temperature limit, where the self-energies exhibit the power-law forms.
 \section{Linearized gap solution for $T> 0$}
\label{Sec:Lin_single_coupling}
The linearized gap equation appears in the main text, Eq. (5). Rewriting this equation we have,
\begin{equation}
    \hat{\Phi}^{\al\bt}_{ij}(\w_n)=\dfrac{g^2_z}{N} T \sum_{m ,\al^\prime, \bt^\prime}  D(\w_n-\w_m) \s^z_{\al\al'} 
   G(\w_m) \hat\Phi^{\al'\bt'}_{ji}(\w_m) G(-\w_m) \s^z_{\bt\bt'} \,.
   \label{SEq:Lin_full_Phi}
\end{equation}
The only attractive channel is $\s_x = -i \s_y \s_z$, as can be verified by substituting various Pauli matrices into $\hat\Phi$ and summing over spin indices. Using the relationships $\s^z \s^x\s^z=-\s^x$, $\hat{\Phi}^{\al\bt}_{ij}=-\hat{\Phi}^{\al\bt}_{ji}$, and $\hat{\Phi}= \s^x\Phi_t \hat{A}$ for the triplet pairing, where $\Phi_t$ and $\hat{A}$ are defined in the main text, we obtain
\begin{equation*}
\Phi_t(\w_n)= \dfrac{1}{N} g_z^2 T \sum_m \dfrac{\Phi_t(\w_m)}{(\w_m+\Sg(\w_m))^2} D(\w_n-\w_m) \,.
\label{Eq:phi_Tnz}
\end{equation*}
We see that the choice of $\hat{A}$ is inconsequential in determining the above equation from Eq.~\eqref{SEq:Lin_full_Phi}. This leads us to the conclusion that $T_c$ for the triplet SC is the same irrespective of $\hat{A}$. For convenience, we define the SC gap function as $\Delta_t(\w_n)=\w_n\Phi(\w_n)/[\w_n+\Sg(\w_n)]$, and using the above equation we derive:
\begin{equation}
\Delta_t(\w_n) = g_z^2 T  \sum_m \dfrac{D(\w_n-\w_m)}{\w_n+\Sg(\w_n)} \Bigg[ \dfrac{1}{N}  \dfrac{\Delta_t(\w_m)}{\w_m} - \dfrac{\Delta_t(\w_n)}{\w_n} \Bigg] \,.
\label{SEq::Lin_gap_eq}
\end{equation}
Comparing this equation with Eq.~(10) in Ref.~\cite{Andrew2021}, we see an equivalence between the parameter $1/N$ and $(1-\al)$, where $\al$ in Ref.~\cite{Andrew2021} determines the strength of time-reversal symmetry breaking disorder. For $\alpha=0$, indicating time-reversal symmetry, random couplings are drawn from the GOE. Conversely, a nonzero $\alpha$ breaks time-reversal symmetry, manifesting as an imaginary part proportional to $\alpha$ in the coupling. Specifically, the random coupling matrix reflects the GUE for $\alpha=1$. We stress however, that while the gap \emph{equation} is equivalent for our model and that of Ref. \cite{Andrew2021}, the gap \emph{structures} are completely different.

A solution of the gap equation has been shown to exist only for $\al$ below a critical value $\al_c$ dependent on $\eta$. 
This suggests a critical value $N=N_c$ in our case as well, below which the SC phase emerges. $N=1$, indicating $\alpha=0$,  corresponds to the largest critical temperature. Moreover, the rigorous validation of our theory requires accessibility of a large $N$ limit, hence a large $N_c$ for the SC phase. As $N_c\rr \infty$, corresponding to $\al\rr 1$, it associates with $\eta\rr 0$, which is achievable if $M\gg N$, indicating the presence of a large number of bosons for each fermion. This is also consistent with the constraint $\sqrt{2M}>N$ for the SC phase to occur.

\begin{figure}
\centering
\includegraphics[scale=0.6]{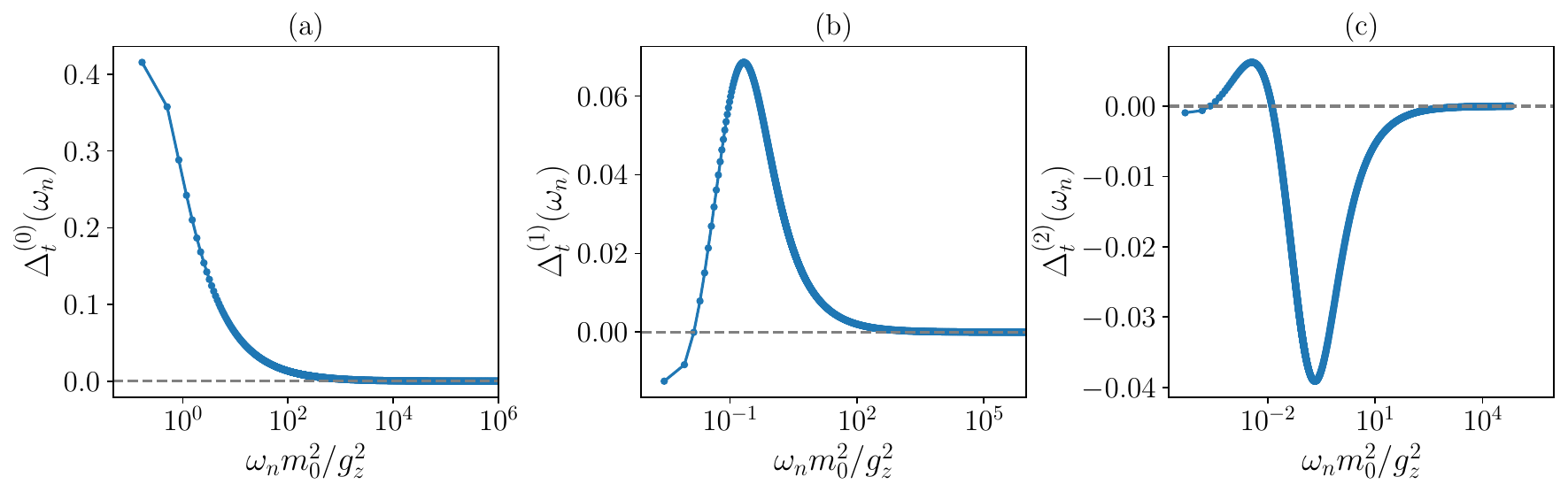
}
\caption{ The solutions $\Delta_t^{(i)}$ as a function of Matsubara frequency $\w_n$ are determined using Eq.~\ref{SEq::Lin_gap_eq} for $T_c^{(0)}=5.372\times 10^{-2}$, $T_c^{(1)}=0.891\times 10^{-3}$, $T_c^{(2)}=1.59\times 10^{-5}$ in (a), (b) and (c), respectively. The remaining parameters, namely $\eta=0.68$, $N=1$, are held constant. For the computation of $\Delta_{t}^{(i)}$ across a wide range of $\w_n$, the hybrid frequency technique detailed in Ref.~\cite{Chubukov_2020_gammaModel} is employed.}
\label{SFig:Sol_lin_gap}
\end{figure}

Eq.~\eqref{SEq::Lin_gap_eq} can be formulated as an eigenvalue problem having multiple eigensolutions. The eigenvectors are denoted as $\Delta_t^{(i)}$, where $i$ indicates the eigenvalue index arranged in descending order. $\Delta_t^{(i)}$ associated with the eigenvalue denoted in Eq.~\eqref{SEq::Lin_gap_eq} is the gap solution and these solutions correspond to distinct critical temperature $T_c^{(i)}$. In the main text, the highest critical temperature $T_c^{(0)}$ is referred to as $T_c$, which follows $T_c\sim g_z^2$. Several solutions are shown in Fig.~\ref{SFig:Sol_lin_gap}. Values of  $T_c^{(i)}$ are given in the unit of $m_0^2/g_z^2$. 
In the context of singlet superconductivity, the significance of solutions with $i > 0$ has been discussed previously in Refs.~\cite{Andrew2021,Chubukov_2021}
and references within. They are not relevant for this work and we do not consider them further here.

All numerical calculations, both here and below, have been performed using the parameters $N=1$ and $\eta=0.68$. Note that, $\eta\approx 0.68$ corresponds to $\al_c\approx 0.63$ as cited in Ref.~\cite{Andrew2021},  resulting in $N_c=\lfloor 2.7\rfloor=2$.

\section{Nonlinear correction in the gap equation for $g_0=0$}  
\label{SSec:NonLinGap}
We derive the nonlinear equation in the presence of a single coupling term. First, let us consider the triplet case, where $g_z\ne 0$ and $g_0= 0$. To visualize the structure of the self-energy and the pairing equation, we express the entire Green's function $\hat{\mt{G}}$ in the Nambu space. In this space, the basis is represented by the following set of operators: 
$$ [c_{1\ur}(\w), c_{1\dr}(\w),.. , c_{N\ur}(\w), c_{N\dr}(\w); c_{1\ur}^\dg(-\w), c_{1\dr}^\dg(-\w),.. ,  c_{N\ur}^\dg(-\w), c_{N\dr}^\dg(-\w)]. 
$$
The matrix
$\hat{\mt{G}}$ is then expressed as
\begin{equation}
\hat{\mt{G}}^{-1}=
\begin{pmatrix}
\hat{G}^{-1}(\w) & \hat{\Phi}\\
\hat{\Phi}^{\dg} & -\hat{G}^{-1}(-\w) \\
\end{pmatrix},
\label{SEq:Nambu_G_brief}
\end{equation}
 where $\hat{G}^{-1}(\w)=G^{-1}(\w)\mathds{1}_{2N\times 2N}$, $G^{-1}(\w)=i(\w+\Sg(\w))$, and $\hat{\Phi}$ represents the matrix for all-to-all pairing. $\hat{\mt{G}}$ can be further expressed in terms of the particle-hole ($G_{ph}$) and particle-particle ($G_{pp}$) components: 
\[
\hat{\mt{G}}=\begin{pmatrix}
\hat{G}_{ph} & \hat{G}_{pp}\\
\hat{G}_{hh} & \hat{G}_{hp} \\
\end{pmatrix} \,.
\]
First, we evaluate the equations for triplet pairing and the derivation for singlet pairing readily follows. The self-energy is given by
\[
\hat{\Sg}(\w_n)=\dfrac{g_z^2}{N}T \sum_{m} D(\w_n-\w_m)\hat{G}_{ph}(\w_m) \,.
\]
As shown above, the normal state Green's function obeys $\hat{\Sg}=\mi\Sg\mathds{1}_{2N\times 2N}$ and 
$\hat{G}_{ph} = G\mathds{1}_{2N\times 2N}$. 
The spin indices are not shown explicitly in the following discussion. 

The pairing equation is given by
\begin{equation}
\hat{\Phi}_{ij}(\w_n)=\dfrac{g_z^2}{N}T \sum_{m} D(\w_n-\w_m)\s^z [\hat{G}_{pp}(\w_m)]_{ji} \s^z \,,
\label{SEq:full_gap_eqn}
\end{equation}
where $\hat{G}_{pp}$ is evaluated using blockwise inversion in Eq.~\eqref{SEq:Nambu_G_brief}: 
\[
\hat{G}_{pp} = 
\hat{G}(\w)\hat{\Phi}\hat{G}_f(-\w),
\] 
where we define 
\[\hat{G}_f (-\w) =[\hat{G}^{-1}(-\w) + \hat{\Phi}^{\dg} \hat{G}(\w) \hat{\Phi}]^{-1}.
\]
Expanding $\hat{G}_f$ for small $\hat{\Phi}$, we get
\[
\hat{G}_f(-\w)
=
\hat{G}(-\w) - \hat{G}(-\w) \hat{\Phi}^{\dg} \hat{G}(\w) \hat{\Phi}  \hat{G}(-\w) + [\hat{G}(-\w) \hat{\Phi}^{\dg} \hat{G}(\w) \hat{\Phi}]^2 \hat{G}(-\w) - ...
\]
The approximation $\hat{G}_f\approx\hat{G}(-\w)$ considered in Eq.~\eqref{SEq:full_gap_eqn} reproduces the linear gap equation. The gap equation, including the higher order term in $\hat{G}_f$, is given as
\begin{equation}
\hat{\Phi}_{ij}(\w_n)=\dfrac{g_z^2}{N} T \sum_{m} D(\w_n-\w_m)\s^z \Big[ G(\w_m)\hat{\Phi} G(-\w_m) - G(\w_m)\hat{\Phi} G(-\w_m) \hat{\Phi}^{\dg} G(\w_m) \hat{\Phi}  G(-\w_m) + ... \Big]_{ji} \s^z \,.
\end{equation}
Using $\hat{\Phi}= \s^x\Phi_t \hat{A}$ for the triplet pairing, where $\Phi$ and $\hat{A}$ are defined in the main text, the above equation yields
\[
\begin{split}
\Phi_t \hat{A} 
&=
-\dfrac{g_z^2}{N} T \sum_{m} D(\w_n-\w_m)\Big[ |G(\w_m)|^{2} \Phi_t \hat{A} - |G(\w_m)|^{4} |\Phi_t|^2 \Phi_t \hat{A}\hat{A}^T\hat{A} + ... \Big]^{\mt{T}_o} \,, \\
&=
\dfrac{g_z^2}{N} T \sum_{m} D(\w_n-\w_m)\Big[ |G(\w_m)|^{2} \Phi_t \hat{A} - |G(\w_m)|^{4} |\Phi_t|^2 \Phi_t \hat{A}\hat{A}^T\hat{A} + ... \Big] \,. \\
\end{split}
\]
$\mt{T}_o$ represents the transpose operator in the orbital basis ($ij\rightarrow ji$) and does not affect the spin, hence $A^{\mt{T}_o}=-A$. It is noteworthy that for the singlet case,  $S$ does not flip sign under $\mt{T}_o$. Consequently, the gap shows attraction in the $\mi\s^y$ channel for the $\s^o$ coupling in the Yukawa term. Multiplying both sides of the above equation with $\hat{A}^T$ and taking trace, we obtain
\[
\begin{split}
\Phi_t  
& \approx
\dfrac{g_z^2}{N} T \sum_{m} D(\w_n-\w_m)\Big[ |G(\w_m)|^{2} \Phi_t - |G(\w_m)|^{4} |\Phi_t|^2 \Phi_t \underbrace{ \dfrac{\Tr(\hat{A}^{T}\hat{A}\hat{A}^{T}\hat{A})}{\Tr(\hat{A}^{T}\hat{A})} }_
{\approx N^2/3}\Big] \,, \\
\Rightarrow
\Phi_t(\w_n)
& \approx
g_z^{2}T  \sum_{\w_m} D(\w_n-\w_m) \Big[  \dfrac{1}{N}\dfrac{1}{|\w_m+\Sg(\w_m)|^2} - \dfrac{N}{3}\dfrac{|\Phi_t(\w_m)|^2 }{|\w_m +\Sg(\w_m)|^4}\Big] \Phi_t(\w_m) \,.
\end{split} 
\]
For the linear term, $\Tr(\hat{A}^{T}\hat{A})$ cancels on both sides of the equation. However, $\Tr(\hat{A}^{T}\hat{A}\hat{A}^{T}\hat{A})/\Tr(\hat{A}^{T}\hat{A})$ depends on $A$, and the consequences of this dependence here and in the mixed state are not studied in this work. For the results shown here, we consider $A$ as given in the main text.

$\Delta_t$ from the above equation follows,
\begin{equation}
\Delta_t(\w_n)= g_z^2 T \sum_m \dfrac{D(\w_n-\w_m)}{\w_m+\Sg(\w_m)} \Bigg[ \dfrac{1}{N}  \dfrac{\Delta_t(\w_m)}{\w_m} 
-\dfrac{N}{3} \dfrac{\Delta^3_t(\w_m)}{\w^3_m}
- \dfrac{\Delta_t(\w_n)}{\w_n} \Bigg] \,.
\label{SEq:Nonlin_sz_gap}
\end{equation}
A similar equation can be obtained for the singlet gap function, where $g_z=0$ and $g_0\ne 0$, and the equation is given by
\begin{equation}
\Delta_s(\w_n)= g_0^2 T \sum_m \dfrac{D(\w_n-\w_m)}{\w_m+\Sg(\w_m)} \Bigg[ \dfrac{1}{N}  \dfrac{\Delta_s(\w_m)}{\w_m} 
-N \dfrac{\Delta^3_s(\w_m)}{\w^3_m}
- \dfrac{\Delta_s(\w_n)}{\w_n} \Bigg] \,.
\label{SEq:Nonlin_s0_gap}
\end{equation}
The triplet and singlet gap equations differ by a factor of $1/3$ in the second term on the right-hand side, arising from $\Tr(\hat{A}^{T}\hat{A}\hat{A}^{T}\hat{A})\approx \Tr(\hat{S}^{4})/3\sim N^4$ for large $N$. Consequently, as the temperature decreases below $T_c$, the triplet gap function grows faster than the singlet one. Solutions of the two equations are related as 
\begin{equation}
\Delta_s(\w_n)=\dfrac{1}{\sqrt{3}}\Delta_t(\w_n) \,,
\label{SEq:Ds_Dt_relation}
\end{equation}
with the corresponding interchange between $g_0^2$ and $g_z^2$.
\section{The gap equation in the presence of both $\s^{z}$ and $\s^{0}$}
Here, we derive the gap equation in the presence of both coupling terms.
Following the derivation in Sec.~\ref{SSec:NonLinGap}, the gap equation is expressed as
\begin{equation}
\hat{\Phi}=\sum_{m}\dfrac{T}{N} D(\w_n-\w_m) |G(\w_m)|^{-2}
\sum_{a=z,0}  g_{\lmb}^2 \s^{a} \Big[ \hat{\Phi}- |G(\w_m)|^{-2} \hat{\Phi} \hat{\Phi}^{\dg} \hat{\Phi} \Big]^{T_o} \s^a \,,
\label{SEq:Gap_trip_sing}
\end{equation}
where $\hat{\Phi}= \Phi_t \hat{A}\s^x + \Phi_s \hat{S}\mi\s^y$, as described in the main text. We define $\tilde{\Sg}(\w_m)=\w_m+\Sg(\w_m)$.
To obtain the equation for the triplet pairing function, we take trace after multiplying both sides of the above equation by $\hat{A}^T$. This yields
\begin{equation}
\Phi_t(\w_n)=
(g_z^2-g_0^2) T \sum_{m} D(\w_n-\w_m)\Bigg[ \dfrac{1}{N} \dfrac{\Phi_t}{|\tilde{\Sg}(\w_m)|^{2}} - \dfrac{1}{|\tilde{\Sg}(\w_m)|^{4}} \Big( \dfrac{N}{3} |\Phi_t|^2 \Phi_t + \dfrac{2N}{3} |\Phi_s|^2 \Phi_t 
+ \dfrac{2}{3} (\Phi_s)^2 \Phi^*_t \Big)
 \Bigg] \,.
 \label{SEq:phi_t_NonLin}
\end{equation}
Similarly, for the singlet gap, we multiply both sides of Eq.~\eqref{SEq:Gap_trip_sing}  by $\hat{S}^T$ and take trace,  resulting in
\begin{equation}
\Phi_s(\w_n)=
(g_0^2-g_z^2) T \sum_{m} D(\w_n-\w_m)\Bigg[ \dfrac{1}{N} \dfrac{\Phi_s}{|\tilde{\Sg}(\w_m)|^{2}} - \dfrac{1}{|\tilde{\Sg}(\w_m)|^{4}} \Big( N |\Phi_s|^2 \Phi_s + \dfrac{2N}{3} |\Phi_t|^2 \Phi_s 
+ \dfrac{2}{3} (\Phi_t)^2 \Phi^*_s \Big)
 \Bigg] \,.
 \label{SEq:phi_s_NonLin}
\end{equation}
In the derivation, we use the relations: $\Tr[\hat{A}^T\hat{A}]\sim N^2 $, $\Tr[\hat{A}^T \hat{A} \hat{A}^T \hat{A}] \approx N^4/3$, $\Tr[\hat{A}^T \hat{A} \hat{S} \hat{S}]\approx N^4/3$,  $\Tr[\hat{A}^T \hat{S} \hat{A}^T \hat{S}]\approx 2N^3/3$, 
 $\Tr[ \hat{S}^T \hat{S}]\approx N^2$,  $\Tr[\hat{S} \hat{A}]=0$, and $\Tr[\hat{S}^4]\approx N^4$ for the large-$N$ approximation. 

 It is evident that the singlet and triplet gap equations are equivalent, except for the prefactor, where the difference in disorder strengths exhibits opposite signs in the two cases, and the factor of $1/3$ present in the first nonlinear term only in the triplet equation. Upon rescaling $\Phi_{s(t)}\rightarrow \Phi_{s(t)}/N$ in Eqs.~\eqref{SEq:phi_t_NonLin} and \eqref{SEq:phi_s_NonLin}, we recover Eq.~(6) in the main text.

The last terms in Eqs.~\eqref{SEq:phi_t_NonLin} and \eqref{SEq:phi_s_NonLin} serve to fix the relative phase between $\Phi_s$ and $\Phi_t$ when both pairings are present in the SC state. The phase difference between $\Phi_t$ and $\Phi_s$ is either $0$ or $\pi$. To see this, consider the system in a singlet pairing state, i.e. $g_0^2>g_z^2$, with temperature just below to $T_c$. Here, $\Phi_s$ can be assumed to be real. Upon lowering the temperature, a nonzero triplet pairing $\Phi_t$ can be induced. The last term in
Eq.~\eqref{SEq:phi_t_NonLin} acts as an attractive potential in the triplet channel if $(\Phi_s)^2 \Phi^*_t$ remains positive. Since $\Phi_s$ is real, this implies $\Phi_t$ is real as well, fixing the relative phase up to $\pi$.
Note that the induced phase $\Phi_t$ also has an out-of-phase feedback effect on $\Phi_s$, but this is a higher-order effect that we neglect here. In principle, deep in the mixed phase, the feedback and induced terms may generate a nontrivial phase relation between $\Phi_s$ and $\Phi_t$.
\subsection{Linearized equation and evaluation of $T_c$}
While at the linear approximation, the two gap equations are completely decoupled, the results in the presence of either of the couplings [Sec.~\ref{Sec:Lin_single_coupling}] cannot be extrapolated here just by replacing, e.g., $g_z^2$ with $g_z^2-g_0^2$ for the triplet gap function. To see this, we consider Eq.~\eqref{SEq:phi_t_NonLin} which after linearization yields
\begin{equation}
\Phi_t(\w_n)=
(g_z^2-g_0^2)  \dfrac{1}{N} T \sum_{m} D(\w_n-\w_m) \dfrac{\Phi_t}{|\tilde{\Sg}(\w_m)|^{2}}
 \,.
 \label{SEq:}
\end{equation}
The effective coupling in the gap equation is $\lmb=g_z^2-g_0^2$, whereas that for the self-energies is $\gad=g_z^2+g_0^2$. Consequently, the corresponding equation for $\Delta_t$ is expressed as
\begin{equation}
\Delta_t(\w_n)= \gad T \sum_m \dfrac{D(\w_n-\w_m)}{\w_m+\Sg(\w_m)} \Bigg[ \dfrac{\lmb}{\gad}\dfrac{1}{N} \dfrac{\Delta_t(\w_m)}{\w_m} 
- \dfrac{\Delta_t(\w_n)}{\w_n} \Bigg] \,.
\label{SEq:Nonlin_gap_both_coupling}
\end{equation}
We define
$$
A_{nm}=\dfrac{\gad}{\pi} \dfrac{D(\w_n-\w_m)}{\tilde{\Sg}(\w_m)} \,,
$$
rescale $T\rightarrow T m_0^2/\gad$ \cite{Andrew2021}, and rewrite Eq.~\eqref{SEq:Nonlin_gap_both_coupling} as
\begin{equation}
\Rightarrow.
\sum_m  \dfrac{1}{(1+Q_n)}  \dfrac{A_{nm}}{2m+1} \Delta_m =  N \dfrac{\gad}{\lmb} \Delta_n \,,
\end{equation}
where, $Q_n=\sum_m A_{nm}/(2n+1)$. We vary $T$ to satisfy the above equation for different values of $0<(\lmb/\gad)<1$ and the temperature corresponds to $T_c$, while $N$ is held at a constant value of 1. Without loss of generality, we consider $\gad=1$ and $0<\lmb<1$. As discussed previously, a comparison with Ref.~\cite{Andrew2021} ensures a critical value $\lmb_{c}$ for a given $N$ and $\eta$, above which the SC phase emerges.

Far away from $\lmb_c$ the difference between $\lambda\sim g_z^2$ and $\gad\sim \lambda$ can be neglected, and $T_c \sim \lambda \sim g_z^2$. Near the transition, the critical behavior is 
$T_c\sim \mathrm{exp}\big[-\pi/\sqrt{\lmb-\lmb_{c} } \big]$~\cite{Andrew2021}. For the parameter $N=1$ and $\eta=0.68$, we find $\lmb_{c}\approx 0.37$.

Note that at the linear level, the triplet and singlet gap equations are dual to each other. Therefore, a similar dependency of $T_c$ for the singlet phase holds true as a function of $|\lmb|=(g_0^2-g_z^2)$ as well.
\subsection{Gap function below $T_c$}
\begin{figure}
\centering
\includegraphics[scale=0.46]{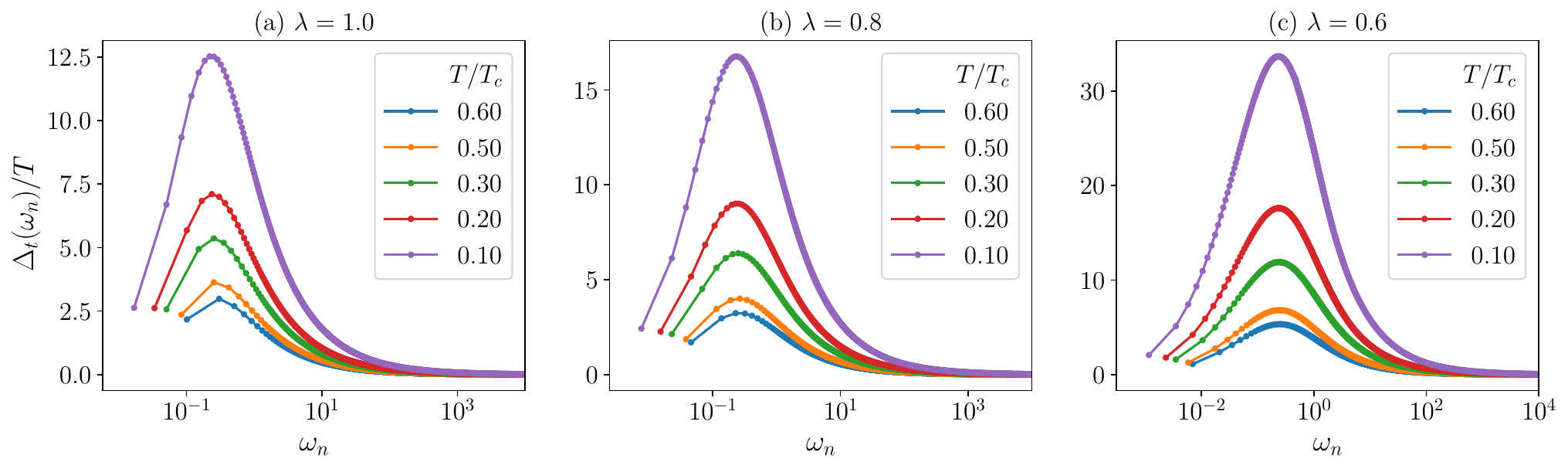}
\caption{The solutions of $\Delta_t(\w_n)$ obtained from Eq.~\eqref{SEq:Nonlin_deltat_both_coupling} for several temperatures $T<T_c$ and coupling $\lmb$. }
\label{SFig:NonLinSol_triplet_both_coupling}
\end{figure}

We aim to obtain $\Delta_t$ for temperatures lesser than but close to $T_c$ and $g_z^2>g_0^2$. In this region, we assume the singlet gap function $\Phi_s=\Delta_s=0$ and verify this assumption later. By using Eq.~\eqref{SEq:phi_t_NonLin}, we derive the nonlinear equation for $\Delta_t$ as
\begin{equation}
\tilde\Delta_t(\w_n)
=
 \dfrac{1}{(1+Q_n)} \dfrac{\lmb}{\gad} \sum_m \Big[\dfrac{1}{N} \dfrac{A_{nm}}{2m+1} \tilde\Delta_t(\w_m) - \dfrac{N }{3\pi^2} \sum_m \dfrac{A_{nm}}{(2m+1)^3}\tilde\Delta_t^3(\w_m) \Big] \,,
 \label{SEq:Nonlin_deltat_both_coupling}
\end{equation}
where $\tilde\Delta_t=\Delta_t/T$ and we set $\gad=1$ as before.
To solve Eq.~\eqref{SEq:Nonlin_deltat_both_coupling}, we use the linear solution $\Delta_{t}^{(0)}(\w_n)$ as the initial input and iterate the equation until convergence is achieved. Several gap solutions for different $T$ and $\lmb$ are shown in Fig.~\ref{SFig:NonLinSol_triplet_both_coupling}. The solution of $\Delta_s$ for $g^2_0>g^2_z$ can be inferred using Eq.~\eqref{SEq:Ds_Dt_relation}.
\begin{figure}
\centering
\includegraphics[scale=0.5]{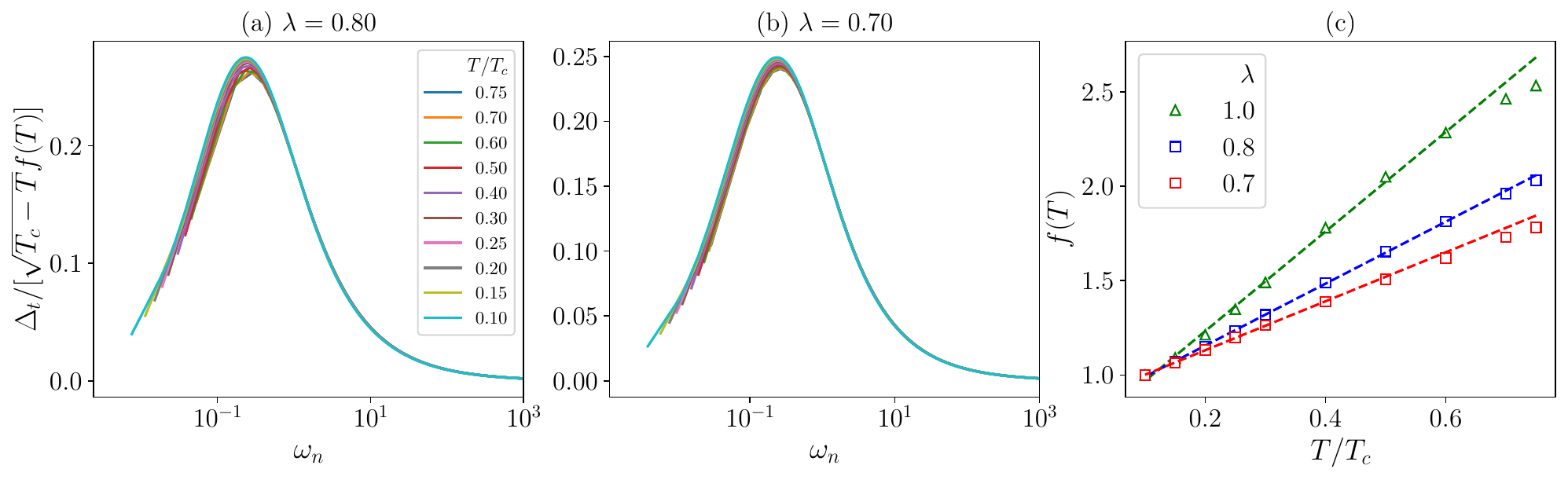}
\caption{\textbf{(a)} and \textbf{(b)}: Various solutions corresponding to different temperatures shown in Figs.~\ref{SFig:NonLinSol_triplet_both_coupling}(b) and (c) are collapsed onto their respective curves corresponding to $T=0.1T_c$ using $f(T)$ as a fitting parameter. \textbf{(c)} The empty markers show $f(T)$ obtained from the collapse. The dashed lines represent the fit of $f(T)\simeq p_1 T/T_c+p_2$ for low temperatures with $p_1$ and $p_2$ given in Tab.~\ref{Stab:fT_params}.}
\label{SFig:FitDt_finiteT_both_coupling}
\end{figure}

Our Ginzburg-Landau expression is formally valid only for $|T_c - T| \ll T$. However, in order to obtain the qualitative form of the phase diagram, we solve it for low $T$ as well. To derive an expression for $\Delta_t$ at low $T$, the finite temperature solutions, such as those represented in Fig.~\ref{SFig:NonLinSol_triplet_both_coupling} are scaled as $\bar{\Delta}_t(T)=\Delta_t/\sqrt{T_c-T}$ and then collapsed onto the lowest temperature curve obtained numerically, denoted as $\bar{\Delta}_{ref}=\bar{\Delta}_t(T=0.1Tc)$ for each value of $\lmb$. The expression used is $
\bar\Delta_{t}(\w_n)\sim f(T,\lmb)\bar\Delta_{ref}
$, where $f(T,\lmb)$ is a fitting parameter. The fittings are depicted in Figs.~\ref{SFig:FitDt_finiteT_both_coupling}(a) and (b) for $\lmb=0.80$ and $0.70$, respectively. Figure~\ref{SFig:FitDt_finiteT_both_coupling}(c) illustrates the functional form of $f(T,\lmb)$, which agrees well with the following analytic form:
\begin{equation}
f(T,\lmb)\simeq p_1(\lmb) \dfrac{T}{T_c}+p_2(\lmb) \,,
\label{SEq:fT_fit}
\end{equation}
for low temperatures with $p_1$ and $p_2$ as the fitting parameters. Values of $p_1$ and $p_2$ are given in Tab.~\ref{Stab:fT_params}. An examination of Eq.~\eqref{SEq:fT_fit} as well as Figs.~\ref{SFig:NonLinSol_triplet_both_coupling} and \ref{SFig:FitDt_finiteT_both_coupling} reveals that the gap \emph{decreases} at low temperatures at the lowest $\w_n$. This is apparently an artifact of the singular form of the expansion at low frequencies, see e.g. Eq.~\eqref{SEq:Nonlin_sz_gap}. Hence, the quantitative form of the phase diagram should not be taken too seriously.
\begin{table}
\begin{center}
\begin{tabular}{ |p{1cm}||p{1cm}|p{1cm}| }
    \hline
         $\lmb$ & $p_1$ & $p_2$ \\
         \hline \hline
        0.55 & 1.24 & 0.91  \\
        \hline
        0.60 & 1.03 & 0.91  \\
        \hline
         0.65 & 1.15 & 0.89 \\
        \hline
         0.70 & 1.30 & 0.87 \\
        \hline
         0.75 & 1.46 & 0.85 \\
        \hline
         0.80 & 1.64 & 0.83 \\
        \hline
         0.85 & 1.84 & 0.80 \\
        \hline
         0.90 & 2.08 & 0.77 \\
        \hline
         0.95 & 2.34 & 0.74 \\
        \hline
          1.0 & 2.64 & 0.71 \\
        \hline
    \end{tabular}
    \caption{Fitting parameters for $f(T,\lmb)$ expressed in Eq.~\eqref{SEq:fT_fit} at several $\lmb$.}
    \label{Stab:fT_params}
\end{center}
\end{table}

\section{Calculation of the critical temperature for the mixed state}
\begin{figure}
\centering
\includegraphics[scale=0.6]{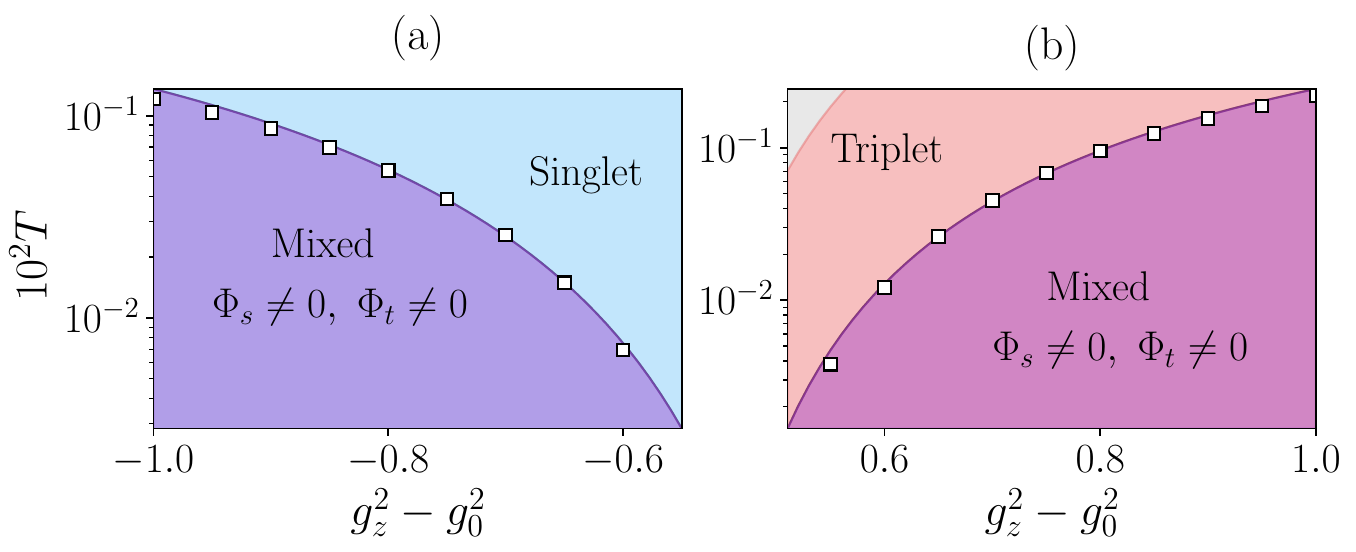}
\caption{Numerically evaluated critical temperatures $T_{s\rr m}$ and $T_{t\rr m}$, denoted by squares, are shown in (a) and (b), respectively. We see that $T_{s\rr m}<T_{t\rr m}$. The solid lines in (a) and (b) illustrate fittings to a curve of the form shown in Eq.~\eqref{SEq:fit_Ttm}.}
\label{SFig:Tmix_both}
\end{figure}

Let us consider the case mentioned in the main text, i.e., $g_0^2<g_z^2$ and $T$ is just below $T_c$, which results in $\Phi_t\ne 0$ and $\Phi_s=0$. The SC phase consists purely of triplet pairing. Now consider lowering the temperature further. We anticipate that $\Phi_s$ will be  small near the  onset of the mixed state and neglect the $\sim\Phi_s^3$ term in Eq.~\eqref{SEq:phi_s_NonLin}, which leads to a linear equation for $\Phi_s$:
\begin{equation}
\Phi_s(\w_n)=
-\lmb \dfrac{T}{N} \sum_{m} D(\w_n-\w_m) \Bigg[ 
\dfrac{\Phi_s}{|\tilde{\Sg}(\w_m)|^{2}}
-
\dfrac{1}{|\tilde{\Sg}(\w_m)|^{4}} \dfrac{2}{3} |\Phi_t|^2 \Phi_s
\Bigg] \,.
\label{SEq:induced_singlet}
\end{equation}
Using the definition of $\Delta_s$, we obtain
\begin{gather}
\tilde{\Sg}(\w_n) \Delta_s(\w_n)=\w_n\Phi_s(\w_n)   \,,\nonumber \\
\Rightarrow
\Delta_s(\w_n)= \gad T  \sum_m \dfrac{D(\w_n-\w_m)}{\tilde{\Sg}(\w_m)} \Bigg[\dfrac{\lmb}{\gad} \dfrac{1}{N} \Bigg( \dfrac{2}{3} \dfrac{\Delta_t^2(\w_m)}{\w_m^2 } - 1 \Bigg) \dfrac{\Delta_s(\w_m)}{\w_m} - \dfrac{\Delta_s(\w_n)}{\w_n} \Bigg] \,.
\end{gather}
This can be written as
\[
\dfrac{\lmb}{\gad} \dfrac{1}{N}\sum_m \dfrac{A_{nm}}{(1+ Q_n) (2m+1)} \Bigg( \dfrac{2}{3} \dfrac{\Delta_t^2(\w_m)}{\w_m^2 } - 1 \Bigg) \Delta_s(\w_m)
=
\Delta_s(\w_n) \,, 
\]
\begin{equation}
\Rightarrow
\dfrac{\lmb}{\gad} \dfrac{1}{N}\sum_m \dfrac{A_{nm}}{(1+ Q_n) (2m+1)} \Bigg( \dfrac{2}{3} \dfrac{\tilde{\Delta}_t^2(\w_m)}{(2m+1)^2 \pi^2 } - 1 \Bigg) \tilde{\Delta}_s(\w_m)
=
\tilde{\Delta}_s(\w_n) \,.
\label{Eq:Induced_triplet}
\end{equation}
Using the finite temperature form of $\Delta_t$ discussed in the previous section,
we solve Eq.~\eqref{Eq:Induced_triplet} as an eigenvalue problem. The critical temperature $T_{t\rr m}$ for the mixed state is obtained as a function of $\lmb$, as shown in Fig.~\ref{SFig:Tmix_both}(b). 
We can estimate the critical temperature using Eq.~\eqref{SEq:Estimate_induce}. At low energy $\Delta_t \sim T^{(1+\eta)/2}\Phi_t$ and assuming the kernel to be zero at the critical temperature, we obtain
\begin{equation}
T_{t\rr m}\sim \mathrm{exp}\Big[-\dfrac{\pi}{\sqrt{\lmb-\lmb_{m_1}}}\dfrac{4}{3+\eta} \Big] \,,    
\label{SEq:fit_Ttm}
\end{equation}
where $\Delta_t(T=0)\sim T_c$.  
The numerical data points fit well with the above form as shown in Fig.~\ref{SFig:Tmix_both}(b) for $\lmb_{m_1}\approx \lmb_c$. 

Similarly, if we begin from a pure singlet state, i.e., $g_0^2>g_z^2$, and lower the temperature, due to the slower growth of the singlet gap function [Eq.~\eqref{SEq:Ds_Dt_relation}], the critical temperature for the transition to the mixed state $T_{s\rr m}$ will be smaller than $T_{t\rr m}$. This is verified numerically and shown in Fig.~\ref{SFig:Tmix_both}(a), where a similar functional form as that for $T_{t\rr m}$ is used to fit $T_{s\rr m}$. The critical value of $g_0^2-g_z^2$ is approximately $\lmb_{m_2}\approx -\lmb_{m_1}$ for the mixed state to appear [Fig.~\ref{SFig:Tmix_both}(a)].
Our numerical analysis and the value of $\lmb_{m_1}\approx -\lmb_{m_2}\approx \lmb_c$ suggest that the ground state of the superconductor always consists of mixed pairings. We now discuss this issue and show that the reality is a bit more complex.

Consider the behavior of the kernel in Eq.~\eqref{SEq:induced_singlet} deep in the SC phase and away from $\lmb_c$.
We replace the sum with integration with the UV cutoff $\w_0$, where $\w_0\sim \Sg(\w_0)$. At $\w_n=\pi T$, this gives
\begin{equation}
\Phi_s(T)\sim A \lmb \int_{T}^{\w_0} d\w \Bigg[\dfrac{1}{B\w^{(3-\eta)/2}} |\Phi_t|^2- \dfrac{1}\omega 
\Bigg] \Phi_s
\,,
\label{SEq:Estimate_induce}
\end{equation}
where all $\eta$-dependent constants are incorporated  into $A$ and $B$. Assuming $|\Phi_t|^2$ to be a constant, the kernel simplifies to 
\[\sim
\dfrac{2|\Phi_t|^2}{(1-\eta)BT^{(1-\eta)/2}}-\ln\dfrac{\w_0}{T}, 
\]
which diverges as $T\rr 0$. The implication is that for low enough temperatures, the nonlinear term always overtakes the linear term. As a result, in the SC phase far away from the QCP, at sufficiently low temperatures, the pure pairing state always transits to a mixed state (although formally the Ginzburg-Landau treatment also breaks down at the same scale). However, near $\lmb_c$, the divergence at $T\rr 0$ renders our entire Ginzburg-Landau approach invalid, as can be seen by e.g. the fact that $\Phi_t$ vanishes at low frequencies in the $T\to 0$ limit, see Fig.~\ref{SFig:FitDt_finiteT_both_coupling}. Along the $T=0$ axis, since the pure singlet/triplet pairing phase undergoes a second-order phase transition at $\lmb_c$, the gap function, being infinitesimally small close to $\lmb_c$, cannot dominate over the repulsive linear term for the induced phase. Consequently, the induced gap function remains repulsive and becomes attractive only when at higher coupling strength the nonlinear term dominates, thereby suggesting the existence of the second QCP. To pinpoint $\lmb$ corresponding to this QCP, the full nonlinear equation needs to be solved, which is beyond the scope of the current endeavor.

\subsection{Onsite pairing}
Up to this point, we have not considered intra-orbital pairing in our discussion. To incorporate it, we extend the pairing function as follows: 
\begin{equation}
\hat{\Phi}=\Phi_t \hat{A}\s^x + [\Phi_{nl}\hat{S}+ \Phi_{l}\hat{\mathds{I}}]\mi\s^y \,,
\end{equation}
where diag$(S)=\{0\}$ as before and $\hat{\mathds{I}}$ represents the identity matrix. To distinguish between the local and nonlocal singlet pairing, we introduce $\Phi_l$ for the local pairing and replace the notation $\Phi_s$ with $\Phi_{nl}$ for the nonlocal pairing.

Following the previously mentioned method, we evaluate the equations for the gap functions as: 
\begin{flalign}
\label{SEq:local-1}
\Phi_t(\w_n)=
(g_z^2-g_0^2) \dfrac{T}{N} \sum_{m} \dfrac{D(\w_n-\w_m)}{|\tilde{\Sg}(\w_m)|^{2}}\Bigg[ 
\Phi_t - 
 \dfrac{\Phi_t}{|\tilde{\Sg}(\w_m)|^{2}} 
\Big\{\dfrac{1}{3} |\Phi_t|^2  + \dfrac{2}{3} |\Phi_{nl}|^2 + 2 |\Phi_{l}|^2 + \dfrac{2}{3} (\Phi_{nl}^* \Phi_{l}+\Phi_{nl} \Phi_{l}^*) \Big\} \nonumber \\
- \dfrac{\Phi_t^*}{|\tilde{\Sg}(\w_m)|^{2}} \Big\{
 \dfrac{2}{3N} (\Phi_{nl})^2  
 - (\Phi_{l})^2  
- \dfrac{2}{3} \Phi_{nl} \Phi_{l} \Big\}
\Big)
 \Bigg] \,
\end{flalign}
\begin{flalign}
\Phi_{nl}(\w_n)=
(g_0^2-g_z^2) \dfrac{T}{N}\sum_{m} \dfrac{D(\w_n-\w_m)}{|\tilde{\Sg}(\w_m)|^{2}} \Bigg[ 
\Phi_{nl} - \dfrac{\Phi_{nl}}{|\tilde{\Sg}(\w_m)|^{2}} 
\Big\{ |\Phi_{nl}|^2  + \dfrac{2}{3} |\Phi_{t}|^2  
+ 2 |\Phi_{l}|^2 + \Phi_{nl}^* \Phi_{l}+ \Phi_{nl} \Phi_{l}^*  \Big\} \nonumber \\
- \dfrac{\Phi_{nl}^*}{|\tilde{\Sg}(\w_m)|^{2}} 
\Big\{
\dfrac{2}{3N} (\Phi_{t})^2 + (\Phi_{l})^2 
+  \Phi_{nl} \Phi_{l}  
\Big\}
 \Bigg] \,\label{SEq:local-2}
\end{flalign}
and
\begin{flalign}
\label{SEq:local-3}
\Phi_{l}(\w_n)=
(g_0^2-g_z^2) \dfrac{T}{N} \sum_{m} \dfrac{D(\w_n-\w_m)}{|\tilde{\Sg}(\w_m)|^{2}}\Bigg[ 
\Phi_{l} - \dfrac{\Phi_{l}}{|\tilde{\Sg}(\w_m)|^{2}} 
\Big\{ |\Phi_{l}|^2 + \dfrac{2}{N} |\Phi_{t}|^2+ \dfrac{2}{N} |\Phi_{nl}|^2 \Big\} 
\nonumber\\
 - \dfrac{\Phi_{l}^*}{|\tilde{\Sg}(\w_m)|^{2}}
\Big\{
- \dfrac{1}{N} (\Phi_{t})^2 
+ \dfrac{1}{N} (\Phi_{nl})^2
\Big\}
 \Bigg] .
\end{flalign}
In these equations, we rescale the triplet and nonlocal singlet pairing functions: $\Phi_t\rightarrow \Phi_t/N$, $\Phi_{nl}\rightarrow \Phi_{nl}/N$. 
For complex order parameters, the equations suggest complicated relative phases among them, which are left for future exploration. 
\end{document}